\documentclass[aps,twocolumn,superscriptaddress]{revtex4}
\usepackage{graphics}
\usepackage[dvips]{graphicx}
\usepackage{times}
\usepackage{color}
\begin{document}

\title{Excessive abundance of common resources deters social responsibility}

\author{Xiaojie Chen}
\email{chenx@iiasa.ac.at}
\affiliation{Evolution and Ecology Program, International Institute for Applied Systems Analysis
(IIASA), Schlossplatz 1, A-2361 Laxenburg, Austria}

\author{Matja{\v z} Perc}
\email{matjaz.perc@uni-mb.si}
\affiliation{Faculty of Natural Sciences and Mathematics, University of Maribor, Koro{\v s}ka cesta 160, SI-2000 Maribor, Slovenia}

\begin{abstract}
We study the evolution of cooperation in the collective-risk social dilemma game, where the risk is determined by a collective target that must be reached with individual contributions. All players initially receive endowments from the available amount of common resources. While cooperators contribute part of their endowment to the collective target, defectors do not. If the target is not reached, the endowments of all players are lost. In our model, we introduce a feedback between the amount of common resources and the contributions of cooperators. We show that cooperation can be sustained only if the common resources are preserved but never excessively abound. This, however, requires a delicate balance between the amount of common resources that initially exist, and the amount cooperators contribute to the collective target. Exceeding critical thresholds in either of the two amounts leads to loss of cooperation, and consequently to the depletion of common resources.
\end{abstract}

\maketitle

Ensuring sustainable use of environmental, social, and technological resources is a global challenge \cite{brito_s12}. Stripped of particularities, the problem is essentially that of responsive use of public goods \cite{ostrom_90}. If the public goods are not managed responsibly, the ``tragedy of the commons'' \cite{hardin_g_s68} is unavoidable. The public goods game is traditionally employed as a theoretical model that describes the social dilemma that emerges when individual short-term interests are inherently different from what would be best for the society as a whole \cite{sigmund_10}. Governed by group interactions, the public goods game requires that players decide simultaneously whether they wish to contribute to the common pool, i.e., to cooperate, or not. Regardless of the chosen strategy, each member of the group receives an equal share of the public good after the initial investments are multiplied by a synergy factor that takes into account the added value of collaborative efforts. Evidently, individuals are best off by not contributing anything to the common pool, i.e., by defecting, while the group would be most successful if everybody cooperated.

Although many mechanisms are known that promote the evolution of cooperation in the public goods game \cite{nowak_jtb12, rand_tcs13, perc_jrsi13}, such as punishment \cite{gachter_s08, rockenbach_n09, helbing_ploscb10, sigmund_n10, boyd_s10, hilbe_srep12, raihani_tee12}, reward \cite{hilbe_prsb10, szolnoki_epl10, szolnoki_njp12}, and social diversity \cite{santos_n08, santos_jtb12}, it has recently been argued that the collective-risk social dilemma game might be more appropriate for capturing the essence of several realistic problems concerning the conservation of common resources \cite{milinski_pnas08}. In particular, the collective-risk social dilemma game describes how the failure to reach a declared collective target can have severe long-term consequences. Opting out of carbon emission reduction to harvest short-term economic benefits is a typical example \cite{vasconcelos_ncc13}. The description of the game is as follows. All players within a group are considered to have an initial endowment that comes from the common pool of available resources. While cooperators contribute a fraction of their endowment to the collective target, defectors retain everything for themselves. The risk level is determined by a collective target that should be reached with the contributions of cooperators. If a group fails to reach this target, all members of the group lose their remaining endowments, while otherwise the endowments are retained. The collective-risk social dilemma game thus accounts directly for the depletion of common resources that may result from opting out of cooperation. Recent experimental and theoretical studies have shown that high risk of collective failure raises the chances for coordinated socially responsible actions \cite{wang_j_pre09, santos_pnas11, chen_xj_epl12, chakra_pcbi12, hilbe_pone13, moreira_srep13}.

However, existing modelling studies have assumed that the contributions of cooperators are independent of the available amount of common resources. Here we depart from this traditional setup by introducing a feedback between the amount of common resources and the amount cooperators contribute to the collective target. Our assumption is that in a harsh environment, when common resources are scarce, cooperators are likely to contribute less. On the other hand, if common resources abound, it seems more reasonable to expect larger contributions towards the collective target. Previous theoretical research has already considered various feedbacks between cooperation and the environment \cite{callaway_n02, akiyama_pd02, tanimoto_pd05}, while a thoroughly documented experimental example is the resource competition in populations of yeast \cite{maclean_n06, gore_n09, sanchez_plosb13}. Yeast prefers to use the monosaccharides glucose and fructose as carbon sources. When these sugars are not available, yeast can metabolize alternative carbon sources such as the disaccharide sucrose by producing and secreting the enzyme invertase. The production and secretion of invertase is costly, yet it creates a common resource that can be consumed by the whole population. It has been reported that increasing the amount of glucose available in the media promotes the growth of defectors (yeast that do not production and secretion invertase), thus decreasing the fraction of cooperators at equilibrium and even driving the cooperators to extinction. On the other hand, as the cooperators decreases in frequency, the amount of glucose which sucrose is hydrolyzed to also decreases \cite{gore_n09}. Therefore, there exists a feedback between the density of cooperators and the amount of monosaccharides in the media \cite{sanchez_plosb13}.

We would like to emphasize that our model is not meant to describe a particular experiment, like the resource competition in populations of yeast that we have just described. Instead, we wish to draw on this example and use it as motivation to propose a minimalist model for a proof of principle --- namely that the excessive abundance of common resources may hinder the evolution of cooperation in the collective-risk social dilemma game. Previous models of public goods have ignored resource dynamics in the common pool, and they have also ignored the fact that cooperative behaviours can be influenced by the degree of resources in the common pool. In the proposed model, we therefore assume that there exists a common resource, a non-empty pool, which provides an initial endowment to every player. Subsequently, every individual must decide whether a certain amount of this endowment will be used to refill the pool or not. Refilling the pool is a socially responsible act of cooperation. Importantly, in subsequent rounds, the endowment issued to every player is proportional to the updated amount of the common resource. If the pool is emptying, the endowments will be less and less, and vice versa if the common resource is managed profitably. However, if the resources in the pool become abundant, we impose an upper bound on the endowment, corresponding to the fact that each individual only needs as much resources from the common pool to be completely satisfied \cite{sanchez_plosb13, smaldino_an13}. The question that we aim to answer is how does the described feedback affect the evolution of cooperation \cite{lehmann_jeb06}. We perform simulations of the collective-risk social dilemma game on structured populations. In the next section, we present results that we have obtained on the square lattice, while results for several other interaction networks are summarized in the Supplementary Information that accompanies this paper.

\section*{Results}

\begin{figure}
\centering
\includegraphics[width=8.4cm]{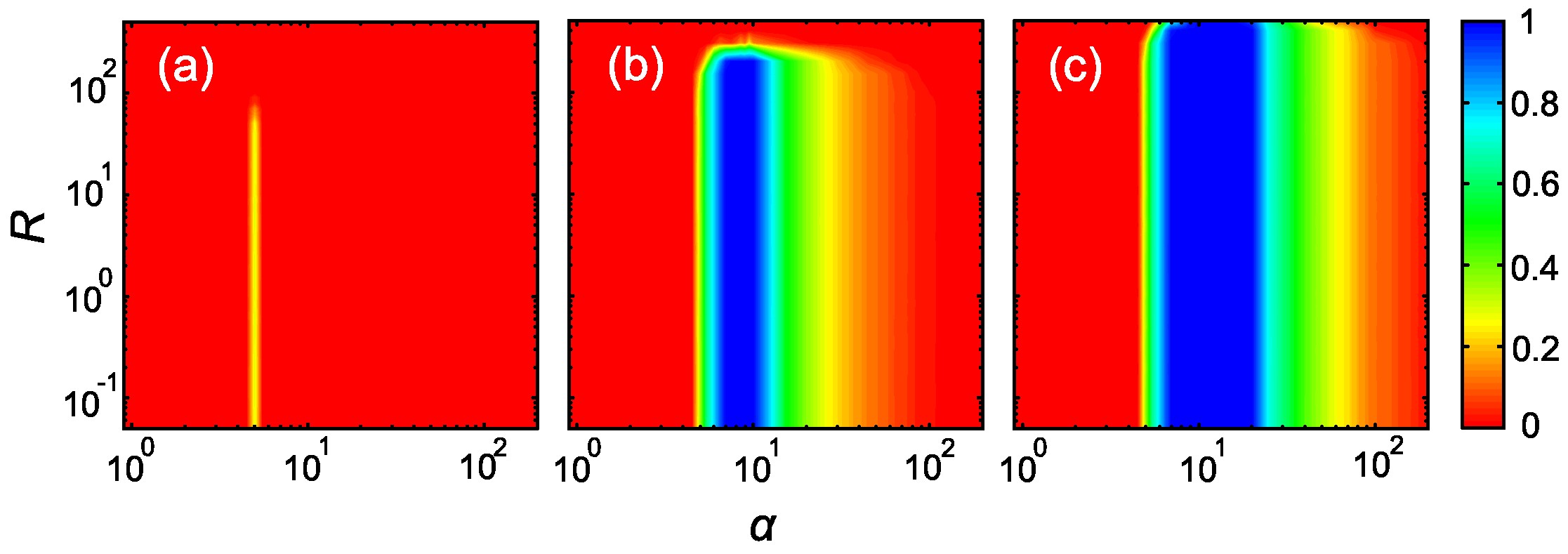}
\caption{Socially responsible actions are viable even if the common resources are initially scarce, as long as the common pool is subsequently kept properly filled. Either too low or too abundant contributions, or failure to distribute them in time, can lead to the tragedy of the commons. Colour maps encode the fraction of cooperators $\rho_c$ in dependence on the multiplication factor $\alpha$ and the initial amount of common resources available to each group $R$, for three different values of the maximal endowment $b$: (a) 5, (b) 10, and (c) 20.}
\label{fig1}
\end{figure}

We begin by presenting colour maps encoding the stationary fraction of cooperators $\rho_c$ in dependence on both the multiplication factor $\alpha$ and the initial amount of common resources $R$ for three different values of the maximal possible endowment $b$. Several interesting conclusions can be drawn from the results presented in Fig.~\ref{fig1}. First, it can be observed at a glance that increasing $b$ (from left to right) increases $\rho_c$ over wide regions of $\alpha$ and $R$. This suggests that if common resources in the population abound, they should be distributed rather than held back. Although this seems to go against cautionary usage and conservation, holding back has, in the long run, several unintended consequences. If the common resources are not distributed right away, defectors can exploit the accumulated stock long after cooperators have disappeared from the neighbourhood. This creates an evolutionary niche for free-riders by means of which they can rise to complete dominance.

The impact of $\alpha$ and $R$ is not as straightforward. As can be inferred from Fig.~\ref{fig1}, only intermediate values of $\alpha$ ensure $\rho_c>0$. However, the span of the optimal interval depends on the maximal endowment $b$. The larger the maximal endowment $b$, the broader the interval of suitable values of $\alpha$. Moreover, there exists an upper bound on $R$, beyond which cooperators cannot survive. The maximal $R$ increases slightly with increasing $b$, but the effect is rather small. Conversely, even if initially the common resources are very scarce, cooperators are not negatively affected provided $\alpha$ and $b$ are from within the limits that ensure $\rho_c>0$. This suggests that cooperative behaviour may develop even under adverse conditions, and it is in fact more likely to do so than under abundance. The extinction of cooperators at both too large $R$ and too large $\alpha$ indicates that an excessive abundance of common resources acts detrimental on the evolution of cooperation, and that thus it deters social responsibility.

\begin{figure}[b]
\centering
\includegraphics[width=7.1cm]{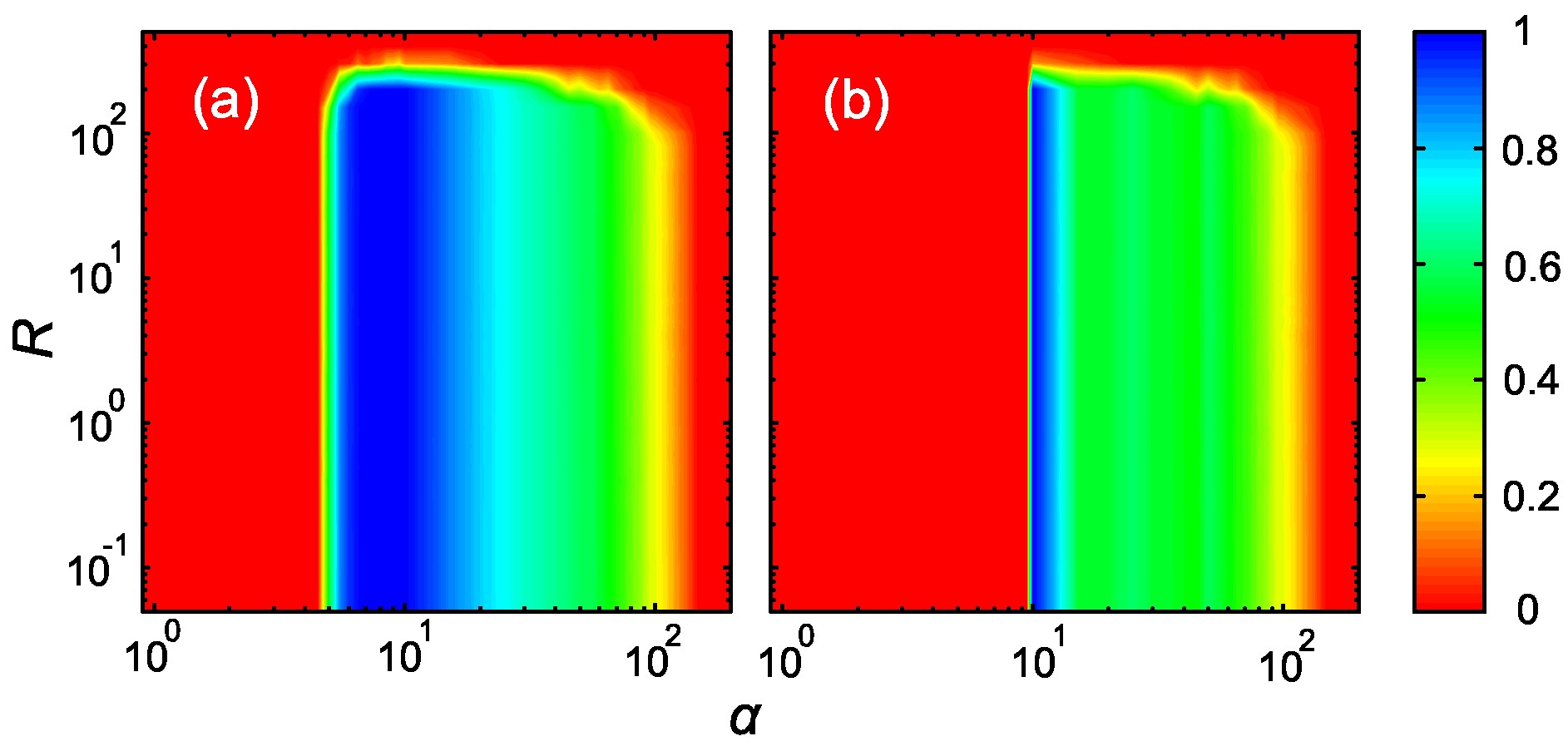}
\caption{Sustainability of common resources is achieved by socially responsible actions. Only an intermediate contribution strength, combined with initially scarce resources, leads to sustainable common resources. In panel (a) the colour map encodes the fraction of groups where the resources can be sustained [i.e., $R_i(\infty)>0$], while in panel (b) the colour map encodes the fraction of groups where the cumulative common goods can provide enough endowments [i.e., $R_i(\infty)\geq Gb$] for all involved. For results in both panels we use the maximal endowment $b=10$.}
\label{fig2}
\end{figure}

To further support our conclusions, we show in Fig.~\ref{fig2}(a) the fraction of groups where the cumulative common goods can be sustained at equilibrium [i.e., $R_i(\infty)>0$], and in Fig.~\ref{fig2}(b) the fraction of groups where the cumulative common goods can provide enough endowments [i.e., $R_i(\infty)\geq Gb$] for all involved. It can be observed that, in comparison to Fig.~\ref{fig1}, the fraction of sustainable groups is larger than zero in a broader region of parameter values. It is much higher than the corresponding fraction of cooperators for large $\alpha$. In combination with Fig.~\ref{fig1}, we thus find that there exists an intermediate region of $\alpha$ that enables cooperators to dominate the population, as well as maintains a sufficient level of common goods in each group for individuals to be fully satisfied. Although the region for such a complete win-win outcome is not broad, it can be broadened by increasing the value of $b$.

\begin{figure}
\centering
\includegraphics[width=8.5cm]{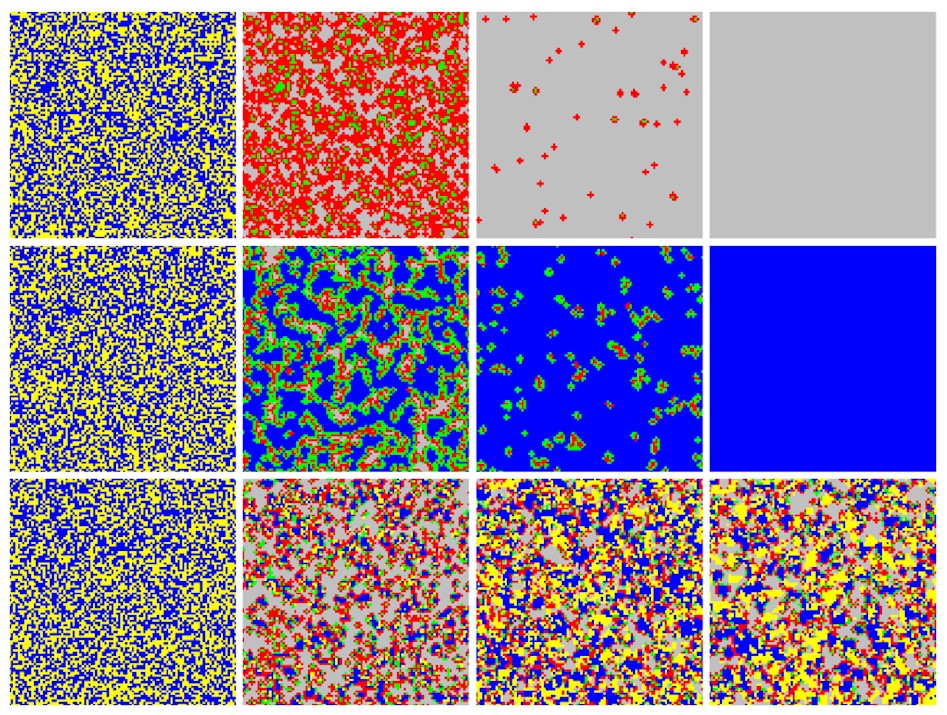}
\caption{Spatial patterns explain why an excessive abundance of
common resources deters social responsibility.
Blue (yellow) are
cooperators (defectors) that are central to groups where the common
resources abound, while green (red) are cooperators (defectors) that
are central to groups where the common resources are scarce. Grey
denotes defectors where the common resources are completely
depleted. Top row show the time evolution (from left to right) for
$\alpha=1$, $b=10$, and $R=50$. Due to the low multiplication factor
the common resources vanish fast. Middle row shows the time
evolution for $\alpha=10$, $b=10$, and $R=50$. Here only cooperative
groups succeed in keeping the pool from emptying. Groups with
defectors quickly become unsustainable and hence pave the way
towards cooperator dominance. Bottom row show the time evolution for
$\alpha=20$, $b=10$, and $R=50$. Due to the high value of $\alpha$
common resources start to abound excessively, making even
predominantly defective groups sustainable and thus fit to invade
cooperators.} \label{fig3}
\end{figure}

The series of snapshots presented in Fig.~\ref{fig3} offers an insight as to what causes the described evolutionary outcomes. We use different colours not just for cooperators and defectors, but also depending on the available amount of common resources. More precisely, blue (yellow) colour denotes cooperators (defectors) that are central to groups where $R_i(t) \geq Gb$. On the other hand, green (red) colour denotes cooperators (defectors) where $R_i(t)<Gb$. Grey are defectors where there are no more common resources left (note that $R_i(t)$ is always larger than zero if cooperators are present). For clarity, we always begin with $R_i(0)=Gb$. Accordingly, blue cooperators and yellow defectors are initially distributed uniformly at random (leftmost panels of Fig.~\ref{fig3}).

For low $\alpha$ (top row of Fig.~\ref{fig3}), the common resources are depleted fast. Defectors turn to red and cooperators turn to green, and widespread grey patches occur only after a few iterations of the game. Soon all is left are isolated islands of defectors who exploit the few remaining cooperators, until eventually all common resources vanish. Consequently, grey defectors come to dominate the entire population. This scenario is characteristic for the case when short-term benefits and ineffective cooperative efforts prevent sustainable management of common resources.

For intermediate $\alpha$ (middle row of Fig.~\ref{fig3}), the scenario is very different. Grouped cooperators are able to preserve and enrich their resources, while groups with defectors fail to do so. Blue cooperative domains, where the common resources abound, become separated from red defectors by strips of green cooperators, which essentially protect the blue domains from being exploited further. The interfaces where green cooperators and red defectors meet become the shield that protects blue cooperative domains. In fact, blue cooperators are able to spread by means of an indirect territorial battle. It is important to note that yellow defectors are practically non-existent, i.e., a defector cannot sustain a profitable group, and accordingly areas of grey soon emerge. These defectors become easy targets once being exposed to blue cooperators.

For high $\alpha$ (bottom row of Fig.~\ref{fig3}), the situation changes again. Here the effectiveness of cooperators is so high that even a few in each group are able to provide more than enough resources for defectors. Accordingly, yellow defectors emerge, which can prevail even against blue domains of cooperators. Note that defectors still have an evolutionary advantage stemming from their refusal to sacrifice a fraction of personal benefits for the conservation of common resources. The stationary state is thus a diverse mix of all possible states, where defectors are more widespread since they don't contribute to the common pool. Nevertheless, if $\alpha$ is not too large some cooperators can still prevail by forming clusters, which as for intermediate values of $\alpha$ are shielded by green cooperators. The ``shield'', however, is not very effective and accordingly has many holes, manifesting rather as isolated green cooperators which signal loss of the blue status rather than forming a compact chain that would prevent the invasion of defectors. If $\alpha$ is larger still (not shown), the utter abundance of common resources leads to the complete dominance of defectors, and ultimately to the tragedy of
the commons as for low values of $\alpha$. The evolutionary path is significantly different though, given that for large $\alpha$ the tragedy is preceded by widespread yellow (rich) rather than red
(poor) defectors. It is also worth noting that large initial values of $R$ result in an identical demise of cooperation as large values of $\alpha$.

To verify the robustness of the presented results, we conclude this section by considering several variations of the proposed collective-risk social dilemma game. In particular, we have studied the effects of (i) the population size, (ii) the topology of the population structure, (iii) different uncertainties by strategy adoptions, (iv) the delay in individual strategy updating, (v) the birth-death update rule \cite{nowak_n04b}, as well as (vi) the effects of cooperator's priority towards limited endowments \cite{gore_n09}. Since the obtained results are not central to the main message of this study, we present all the details and the obtained results in the Supplementary Information. Most importantly, we find that on structured populations our conclusions remain intact under all considered circumstances, thus indicating a high degree of universality. Nevertheless, we emphasize that our conclusions could be challenged under well-mixed conditions. Well-mixing will break up the clusters that we have described in the preceding paragraphs, and this may change the results in favour of non-cooperation. This may be particularly relevant for human cooperation \cite{gracia-lazaro_pnas12, rand_tcs13}, where the movements of and between groups could introduce well-mixed conditions.

\section*{Discussion}

We have shown that an excessive abundance of common resources deters socially responsible actions on structured populations. If either the common resources are initially too many, if the cooperators are too effective in refilling the pool, or if the maximally allowed endowments are too low for allowing an immediate dissemination of accumulated goods, the defectors are able to take full advantage of their refrain from contributing without suffering the consequences. If sufficiently abundant, the excess allows defectors to free-ride well over the time horizon that is required for cooperators to die out. Once this happens the tragedy of the commons cannot be averted. Less surprisingly, if initially the commons resources are absent, or if the efforts of cooperators are ineffective to a degree that the pool becomes empty, the tragedy sets in as well.

The key to sustainability is to properly adjust maximal endowments, which must go hand in hand with the availability of common resources. An abundance of common resources must be felt by all individuals belonging to the group, and the rewards must be administrated fast. This reinforces social responsibility and reimburses cooperators for their preceding selfless efforts. Failure
to do so sooner or later means that the common resources are there for the taking without the need to cooperate. A downward spiral emerges, because the depletion of common resources averts from cooperation also those that previously might have felt it was a viable strategy to adopt. All that is eventually left is a depleted common resource and widespread defection, despite the brief period of excessive abundance.

Based on our findings, as well as based on existing theoretical and experimental research \cite{maclean_n06, andras_bmceb07, gore_n09, requejo_prl12, requejo_pre12, requejo_pre13}, we may conclude that cooperation is the more likely outcome if initially the common resources are scarce rather than abundant. In particular, this conclusion is in agreement with data from experiments conducted on yeast \cite{maclean_n06, gore_n09} as well as on social vertebrates \cite{shen_nc11}. In particular, when the amount of glucose available in the media is increased, defective yeast that do not pay a cost for producing invertase can spread faster than cooperative yeast, even driving cooperative yeast to extinction \cite{maclean_n06, gore_n09}. Similarly, experiments on social vertebrates indicate that unfavourable environmental conditions, where resources are limited, reduce social conflict and make social vertebrates more cooperative \cite{shen_nc11}. We hope that the demonstrated importance of the feedback between cooperative behavior and the availability of common resources will inspire further research aimed at understanding the evolution of cooperation, not least in human societies \cite{rand_tcs13}, where the consideration of mobility might lead to particularly interesting results.

\section*{Methods}
The game is staged on a $L\times L$ square lattice with periodic
boundary conditions. As demonstrated in the Supplementary
Information, changing the topology of the interaction network does
not affect the main conclusions of this study. Each player on site
$x$ with von Neumann neighborhood is a member of five overlapping
groups of size $G=5$, and it is initially designated either as a
cooperator ($s_x=1$) or defector ($s_x=0$) with equal probability.
At time $t$, the endowment $a_x^i$ from group $i$ is defined as
\begin{equation} a_x^i=\left\{
\begin{array}{lll}
b & \mbox{ if }R_i(t)\geq Gb,\\
R_i(t)/G & \mbox{ if } R_i(t)<Gb,
\end{array} \right.
\end{equation}
where $R_i(t)$ is the amount of common resources (public goods)
available to the group at the time, and $b$ determines the maximal
possible endowment an individual is able to receive. As noted
before, this is to take into account that there is only so much an
individual needs \cite{sanchez_plosb13, smaldino_an13}, regardless
of how abundant the common resource may become. Cooperators
contribute a fixed amount $c$ to the common pool in order to prevent
its depletion. Defectors contribute nothing. Accordingly, the payoff
of player $x$ from group $i$ is thus $P_x^i=a_x^i-s_x c$, while the
total payoff $P_x$ is simply the sum over all $P_x^i$ received from
groups where $x$ is a member.

We note that the introduction of a ceiling ($b$) to the endowment is the simplest way by means of which we take into account that, beyond a certain amount, higher endowments will yield no additional returns. Future modelling studies could address more realistic scenarios, for example such where fitness gains continue to increase with increasing endowment but there are diminishing returns. While we do not expect qualitatively different results, the gradual decline of returns with higher endowments might delay the onset of cooperation and affect the parameter values at which we observe the highest levels of cooperative behavior.

Starting with $R_i(0)=R$ in all groups, the amount of common
resources in each group $i$ is updated according to
\begin{equation}
R_i(t+1)=R_i(t)+\sum_{x \in i} (\alpha s_x c - a_x^i),
\end{equation}
where $\alpha$ is the multiplication factor to the amount
cooperators contribute to refilling the pool, thus taking into
account synergetic effects of group efforts. For simplicity, we set
$c=1$, while $b$, $R$ and $\alpha$ are the three key parameters
determining the evolutionary dynamics of the game.

After each round of the game, player $x$ is given the opportunity to
imitate the strategy of one randomly selected nearest neighbour $y$.
The strategy transfer occurs with the probability
\begin{equation}
q=\frac{1}{{1 + \exp [({P_x} - {P_y})/K]}},
\end{equation}
where $K$ is the uncertainty by strategy adoptions
\cite{szabo_pre98}. Without losing generality
\cite{szolnoki_pre09c}, we use $K=0.5$, so that it is very likely
that better performing players will be imitated, although those
performing worse may occasionally be imitated as well.

As the key quantity, we measure the stationary fraction of
cooperators $\rho_c=L^{-2}\sum_x s_x(\infty)$, where $s_x(\infty)$
denotes the strategy of player $x$ when the system reaches dynamical
equilibrium, i.e., when the average cooperation level becomes
time-independent. Moreover, we average the final outcome over $100$
independent initial conditions.

\clearpage

\section*{SUPPLEMENTARY INFORMATION}
\setcounter{figure}{0}

To verify the robustness of the main results presented in the paper, this supplementary information is devoted to the study of several variations of the proposed collective-risk social dilemma. In particular, we study the effects of the population size (\S I), the topology of the population structure (\S II), different uncertainties by strategy adoptions (\S III), the delay in individual strategy updating (\S IV), the birth-death update rule (\S V), as well as the effects of cooperator's priority towards limited endowments (\S VI). In general, our results remain valid under all considered circumstances.

\subsection*{I.\ \ Population size}

\begin{figure}[ht]
\centering
\includegraphics[width=6cm]{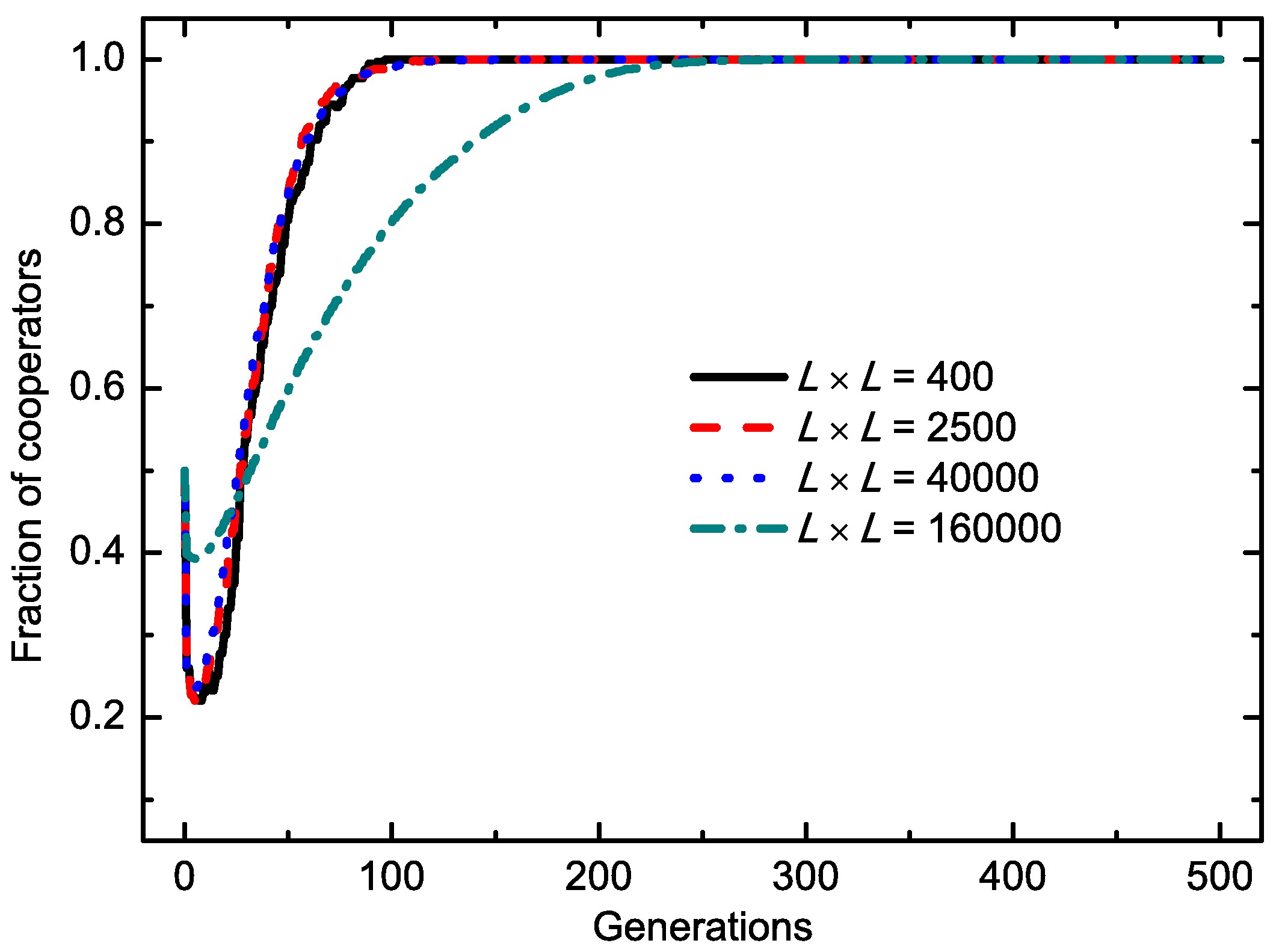}
\caption{Time evolution of the fraction of cooperators for different population sizes (see legend). Parameter values are: $b=10$, $R=10$ and $\alpha=10$.} \label{figs1}
\end{figure}

In this section, we present the time evolution of the fraction of cooperators for different population sizes, to see how the outcome of the evolutionary process depends on this quantity. Figure \ref{figs1} shows that increasing the population size does increase the time for the system to reach the stationary state, but it does not affect the composition of the strategies. For larger population sizes, the system simply needs longer to reach the stationary state.

\subsection*{II.\ \ Population structure}

\begin{figure}
\centering
\includegraphics[width=8.5cm]{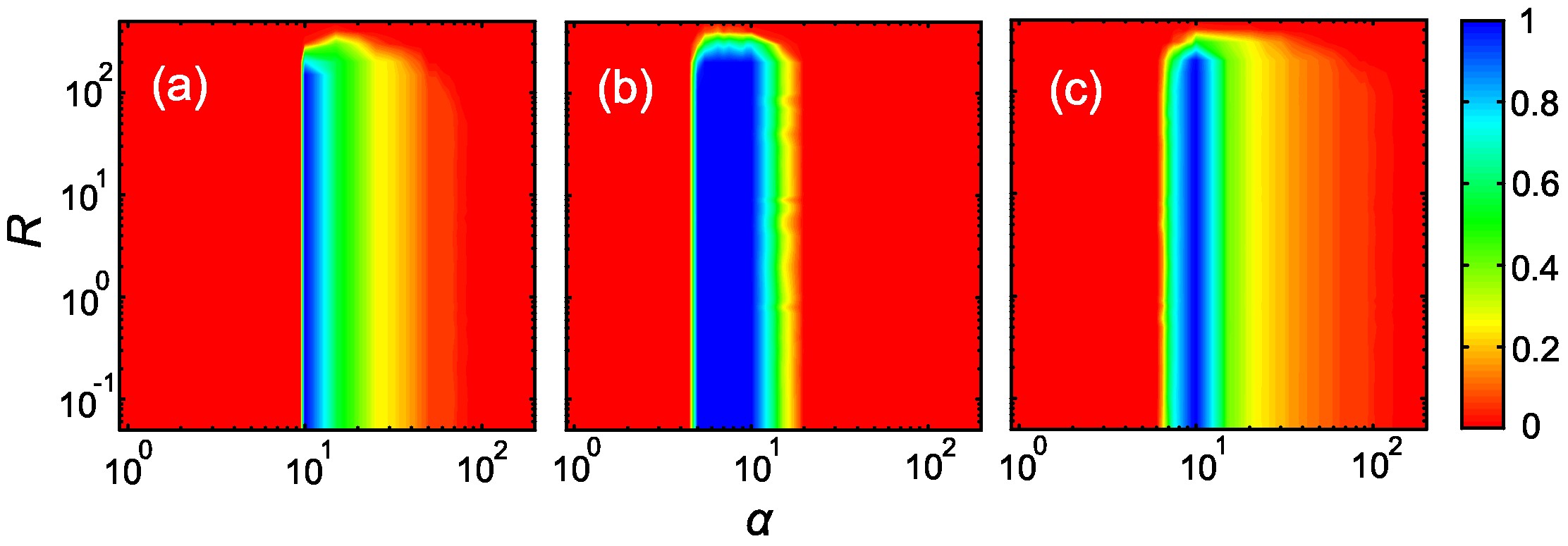}
\caption{The stationary fraction of cooperators in dependence on the multiplication factor and the initial amount of common resources, as obtained for different population structures: (a) random network with group size $G=5$, (b) regular ring with group size $G=5$, and (c)
square lattice with Moore neighborhood ($G=9$). Parameter values are: $b=10$,
$K=0.5$, and the population size is $10^4$ in all three cases.} \label{figs2}
\end{figure}

In this section, we consider the studied collective-risk social dilemma on a random network, on the regular ring, and on the square lattice with Moore neighborhood. On a random network, each individual forms a group with other $G-1$ individuals randomly
chosen from the whole population, and gets its payoff only from the
interactions within the group. While in other interaction networks, each
individual participates in all the $G$ groups that are centered not only on
itself, but also on its nearest neighbors. Figure \ref{figs2} shows the stationary fraction of
cooperators in dependence on $\alpha$ and $R$. It can be observed
that there exist intermediate $\alpha$ values maximizing the
fraction of cooperators, and there exists an upper bound value of
$R$ beyond which cooperators die out. This indicates that our
findings are robust against the changes in the structure of the interaction networks.

\subsection*{III.\ \ Uncertainty by strategy adoptions}

\begin{figure}[b]
\centering
\includegraphics[width=6cm]{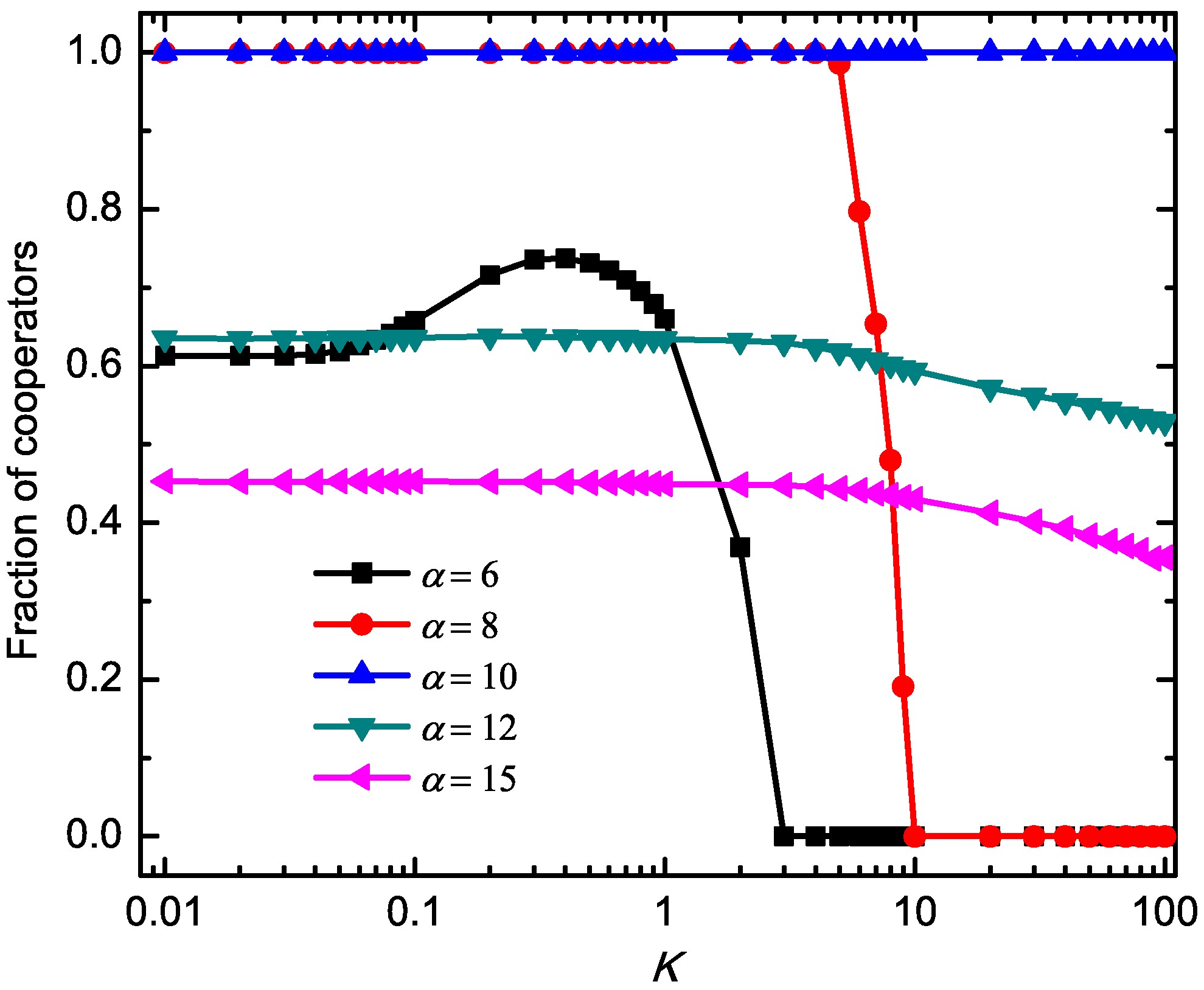}
\caption{The stationary fraction of cooperators in dependence on the
uncertainty parameter $K$ for different values of the multiplication
factor $\alpha$, as obtained on a square lattice with von Neumann neighborhood. Parameter values are: $b=10$ and $L^2=10^4$.}
\label{figs3}
\end{figure}

In this section, we demonstrate the effects of different uncertainty by strategy adoptions on the evolution of cooperation in the studied collective-risk social dilemma. From results presented in Fig.~\ref{figs3} it follows that the stationary fraction of
cooperators varying with $K$ displays four different types of behavior,
depending on the value of the multiplication factor $\alpha$. First,
for a relatively small value of $\alpha$, the fraction of cooperators
first increases slowly until reaching the maximum value, and then
decreases dramatically to zero with increasing $K$. Second, for a
slightly larger $\alpha$ value, full cooperation can be achieved
when $K$ varies from $0.01$ to $5$. But the fraction of cooperators
dramatically decreases to zero with further increasing $K$ from $5$.
Third, for an appropriately intermediate $\alpha$ value, full
cooperation can always be achieved when $K$ varies in a large range
$[0.01, 100]$. Fourth, for a much larger $\alpha$ value, the
fraction of cooperators declines slightly with increasing $K$. This
qualitatively different behavior has been revealed in previous works, and
here we demonstrate again that the uncertainly by strategy adoptions plays
an important role by the evolution of cooperation \cite{SSszabo_pr07, SSperc_jrsi13}.

\subsection*{IV.\ \ Delay in individual strategy updating}

\begin{figure}[ht]
\centering
\includegraphics[width=6.3cm]{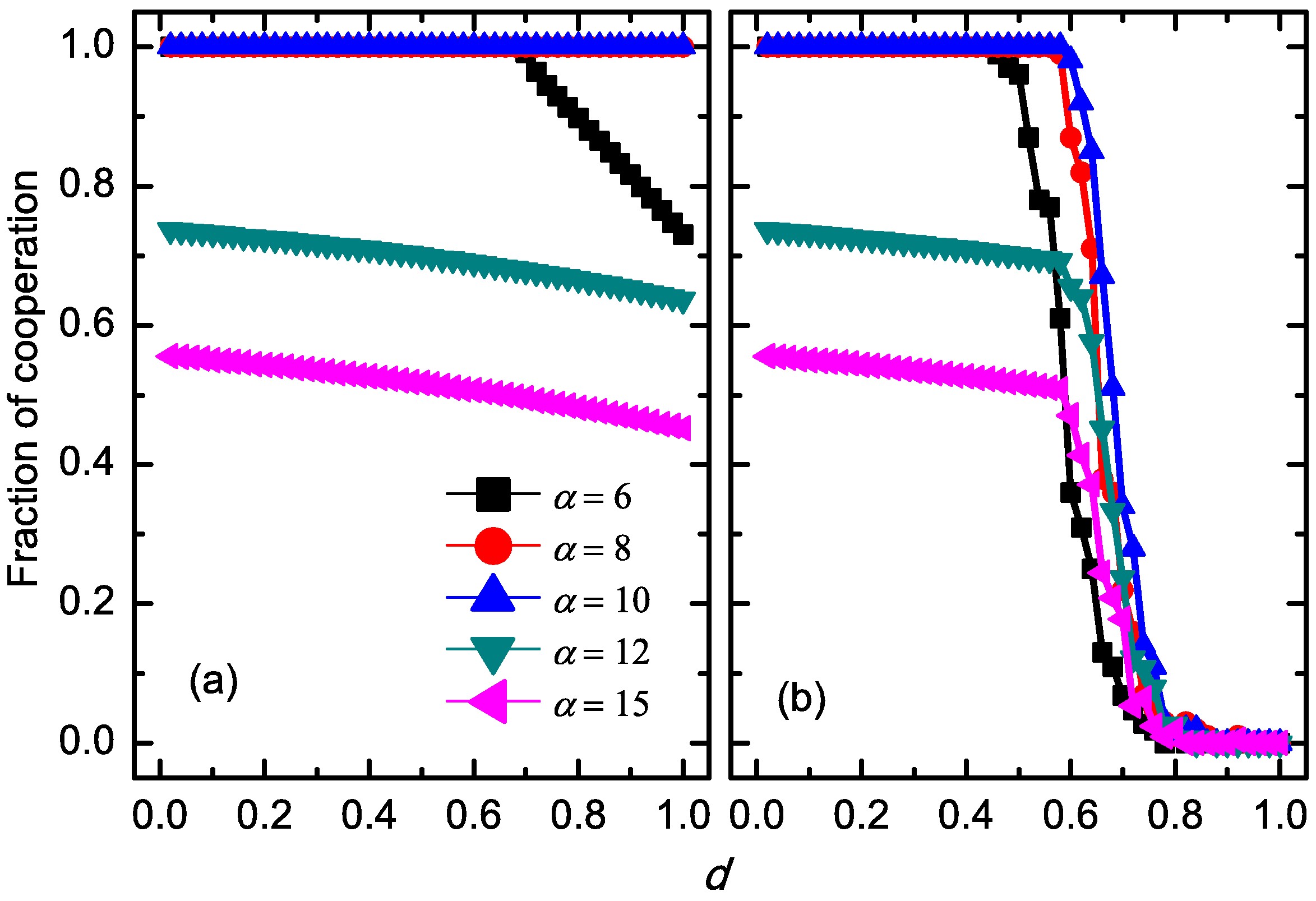}
\caption{The stationary fraction of cooperators in dependence on the
delay parameter for individual strategy update for different values
of the multiplication factor $\alpha$ on a square lattice with von
Neumann neighborhood. Here, $b=10$ and the size of the square lattice is
$L^2=10^4$. The initial amount of common resources $R$ is $10$ in (a)
and $500$ in (b).} \label{figs4}
\end{figure}

In this section, we consider that each individual is subject to delay during strategy updating. While as before, an individual has the opportunity to imitate
the strategy of one randomly chosen neighbor $y$, the probability for this step to be attempted is $d<1$ (rather than $d=1$, as in the main paper). On the other hand, the amount of common resource in each group is still updated at each time step. Figure \ref{figs4}
shows the stationary fraction of cooperators as a function of $d$
for different intermediate values of $\alpha$. For small initial
$R=10$ in panel (a), the fraction of cooperators varying with $d$
displays three different types of behavior, depending on the value of the
multiplication factor $\alpha$. First, for a relatively small
$\alpha$ value, full cooperation is achieved for the delay factor
$d<0.70$. But with increasing $d$ from $0.70$, the fraction of
cooperators decreases quickly. Second, for slightly larger $\alpha$
values, full cooperation can always be achieved, irrespective of the
value of $d$. Third, for much larger $\alpha$ values, the fraction
of cooperators gradually decreases with increasing $d$. In addition,
for much small or much large values of $\alpha$, full defection is
always achieved, irrespective of the value of $d$ (not shown here).
For large initial $R=500$ in panel (b), we see that the fraction of
cooperators first decreases slowly with increasing $d$, and then
dramatically decreases to zero after $d$ reaches a critical value,
and that this is the case for several different intermediate values of $\alpha$.

\subsection*{V.\ \ Birth-death update rule}

\begin{figure}
\centering
\includegraphics[width=3.7cm]{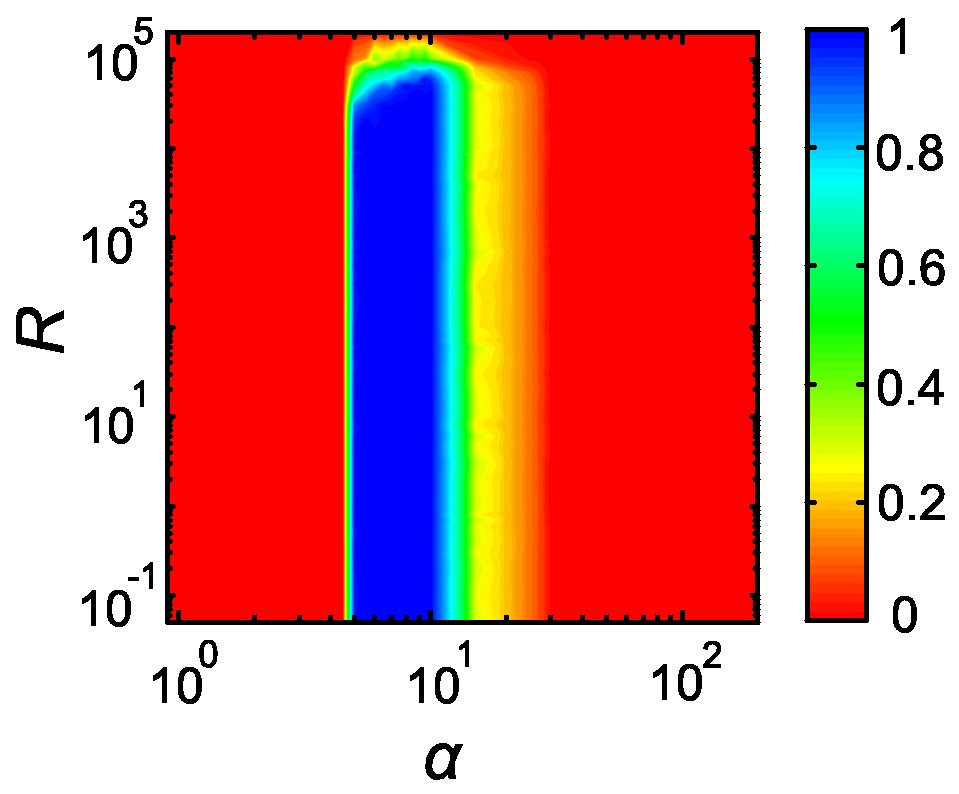}
\caption{The stationary fraction of cooperators in dependence on the
multiplication factor $\alpha$ and the initial amount of common resources $R$, as obtained with the birth-death update rule on a square lattice. Here, $b=10$, $w=0.5$, and the linear population size is $L=20$.} \label{figs5}
\end{figure}

With the motivation to consider the studied collective-risk social dilemma in a perhaps biologically more relevant context, we consider
the birth-death rule instead of the imitation rule used in the main
text. Under the birth-death rule, an individual is chosen for
reproduction proportional to fitness at each time step, and then the
offspring replaces a random neighbour. Because the fitness has to be
positive, following previous works \cite{SStraulsen_bmb08, SSrand_pnas13},
we define an individual $x$'s fitness as $f_x=\exp(wP_x)$, where $w$
$(w>0)$ is the selection intensity. Figure \ref{figs5} depicts that
there exist intermediate $\alpha$ values maximizing the fraction of
cooperators, and there exists an upper bound value of $R$ beyond
which cooperators go extinct. This indicates that our findings are
robust against the changes of update rule. Moreover, the upper bound
value of $R$ is much higher than the one determined under the imitation rule. In fact,
under birth-death update rule at each time step the amount of common
resource in each group is updated, but only one individual is chosen
for updating the strategy. Thus, the amount of common resources is
updated faster than individual strategies, and the upper bound thus becomes
larger.

\subsection*{VI.\ \ Cooperator's priority towards limited endowments}

\begin{figure}[b]
\centering
\includegraphics[width=6cm]{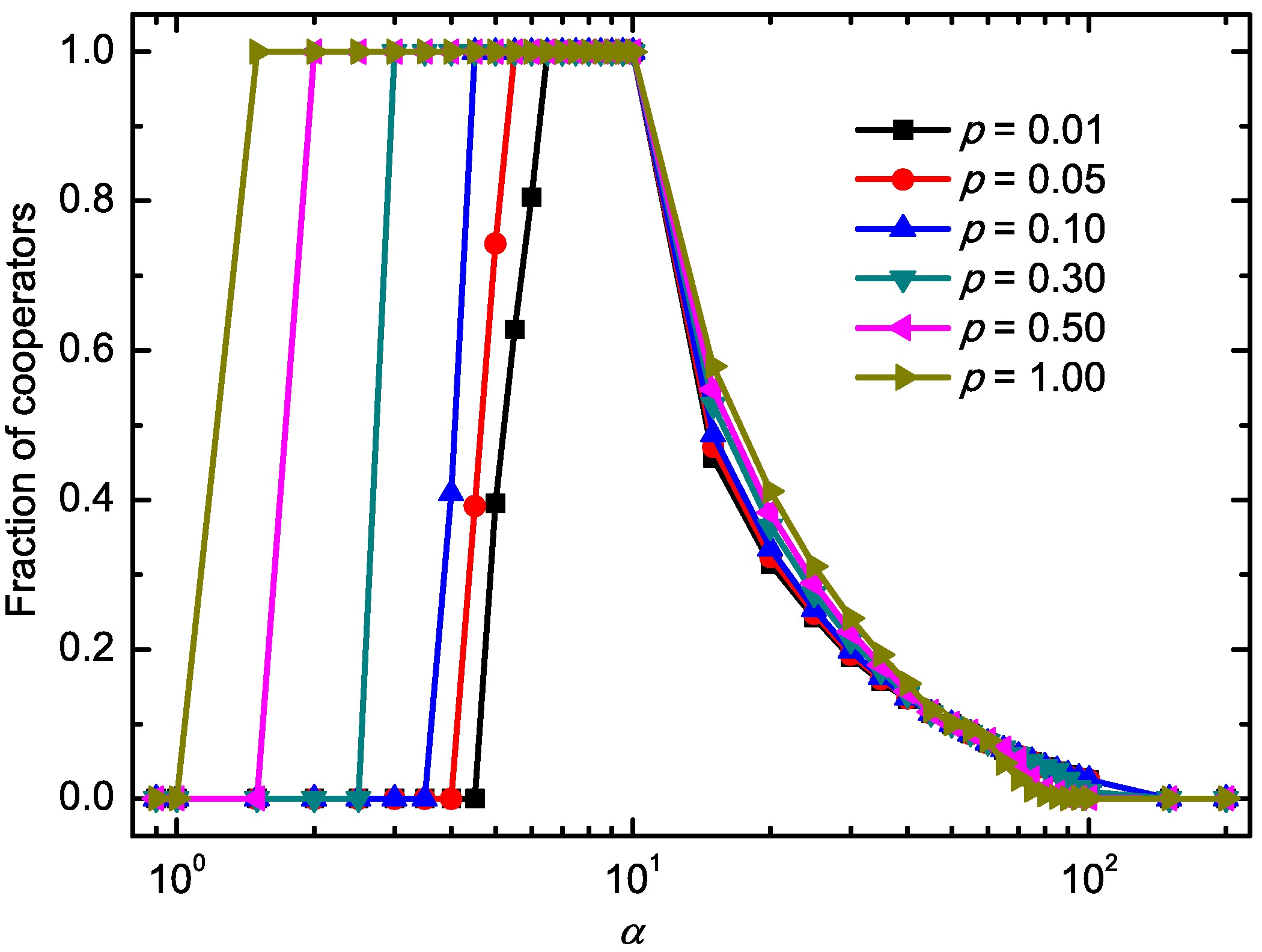}
\caption{The stationary fraction of cooperators in dependence on the
multiplication factor $\alpha$ for different probability values of
cooperator's priority $p$. Other parameter values are: $b=10$, $K=0.5$, $R=10$, and the size of the square lattice is $L^2=10^4$.}
\label{figs6}
\end{figure}

Inspired by Ref.~\cite{SSgore_n09}, in this section we consider that cooperators
can have the priority to use the common resource they produced,
especially when the common resource is limited. To be specific, when
the common resource is not abundant, cooperators can preferentially
to obtain an endowment at a certain probability $p$. Then, a
cooperator's endowment from group $i$ is given as
\begin{equation} a_c^i=\left\{
\begin{array}{lll}
b & \mbox{ if } n_cb\leq R_i(t)<Gb,\\
R_i(t)/n_c & \mbox{ if } 0<R_i(t)<n_cb,
\end{array} \right.
\end{equation}
and a defector's endowment from the same group is given as
\begin{equation} a_d^i=\left\{
\begin{array}{lll}
\frac{R_i(t)-n_cb}{G-n_c} & \mbox{ if } n_cb\leq R_i(t)<Gb,\\
0 & \mbox{ if } 0<R_i(t)<n_cb,
\end{array} \right.
\end{equation}
where $n_c$ is the number of cooperators in group $i$. With
probability $1-p$, an individual's endowment is assigned according
to the method in the main text. In Fig. \ref{figs6}, we show that
there still exists an intermediate value of the multiplication
factor inducing the maximal fraction of cooperators for different
value of $p$, when the initial amount of common resource is limited.
In addition, increasing the $p$ value can further favor the
evolution of cooperation. This is because, if cooperator's priority towards limited common resource is amplified, cooperators simply obtain more opportunities to collect a higher payoff than defectors.

\begin{acknowledgments}
This research was supported by the Slovenian Research Agency (Grant J1-4055).
\end{acknowledgments}

\end{document}